\documentstyle[12pt]{article}
-3\baselineskip \advance\textheight by 4\baselineskip \parskip1.3ex plus0.2ex
minus0.2ex 
\begin{document} \newcommand{\be}{\begin{equation}}
\newcommand{\ee}{\end{equation}} \newcommand{\bea}{\begin{eqnarray}}
\newcommand{\eea}{\end{eqnarray}} \newcommand{\btab}{\begin{tabular}}
\newcommand{\etab}{\end{tabular}} \newcommand{\bef}{\begin{figure}}
\newcommand{\eef}{\end{figure}} \newcommand{\bt}{\begin{table}}
\newcommand{\et}{\end{table}} \newcommand{\ben}{\begin{enumerate}}
\newcommand{\een}{\end{enumerate}} \newcommand{\ba}{\begin{array}}
\newcommand{\ea}{\end{array}} \newcommand{\twowaypartial}{i
\!\!\stackrel{\leftrightarrow}{\partial}} \newcommand{\half}{\frac{1}{2}}
\newcommand{\simleq}{\stackrel{<}{\sim}}
\newcommand{\simgeq}{\stackrel{>}{\sim}} \newcommand{\slashp}{\not \!p}
\newcommand{\slashq}{\not \!q} \newcommand{\slashk}{\not \!k}
\newcommand{\slashepsilon}{\not \!\epsilon} \newcommand{\slashP}{\not
\!\!\!\:P} \newcommand{\slashQ}{\not \!\!\!\:Q} \newcommand{\slashK}{\not
\!\!\!\:K}
 \thispagestyle{empty} \centerline{\begin{tabular}{c} {\Huge Systematic gauge
invariant approach to}\\ {\Huge heavy quarkonium decays}\\
\\
{\bf Hafsa Khan}   \\
Department of Physics    \\
Quaid-e-Azam University  \\
Islamabad, Pakistan. \\
and    \\
{\bf Pervez Hoodbhoy}   \\
Center for Theoretical Physics  \\
Massachussets Institute of Technology   \\
Cambridge, MA 02139, USA   \\
and  \\
Department of Physics    \\
Quaid-e-Azam University  \\
Islamabad, Pakistan.
\end{tabular}                                             }
\pagebreak

\begin{abstract}
We present a method which, starting directly from QCD, permits a systematic
gauge-invariant expansion to be made for all hard processes involving
quarkonia in powers of the quark relative velocity, a small natural parameter
 for heavy quark systems.
Our treatment automatically introduces soft gluons in the expansion.
Corrections arising from the incorporation of gauge symmetry turn out to be
 important for decay and fragmentation processes involving $Q \bar{Q}$
systems. The contribution of soft gluons is shown to be of higher order
in $v$ and so is neglected for calculations done upto and including O($v^2$).
\end{abstract}

\section*{INTRODUCTION}

The principles of quantum chromodynamics (QCD) were applied almost 20 years
ago to the bound states of heavy quarks, such as $c \bar{c}$ and $b \bar{b}$.
 These are possibly the simplest strongly bound systems that exist. The large
 mass of the heavy quark sets a mass scale large enough so that perturbative
QCD, together with a non-relativistic potential model description of the bound
 state, provides a good starting point to describe the decay and formation of
 quarkonia. However, quantitative predictions of the simple quarkonium model,
even supplemented by radiative corrections, sometimes fail badly
\footnote{For a recent review of quarkonium phenomenology see, for example,
Schuler$^{(1)}$.}. Over the years, hundreds of papers have been written to
rectify some of the failures. Nevertheless one still does not have a complete
 solution to this important problem of non-perturbative QCD.

Our investigation into this venerable subject was prompted by the observation
that the fundamental symmetry to which QCD owes its origin, gauge symmetry,
is manifestly violated by the naive quarkonium model. This is not hard to see:
 under a local gauge transformation
$q(\vec{x},t) \longrightarrow U(\vec{x},t) q(\vec{x},t)$
the state normally used to describe quarkonia,
\begin{equation}
\int d^3x_1 d^3x_2 f(\vec{x_1} - \vec{x_2}) \bar{q}(\vec{x_1},t) \Gamma
q(\vec{x_2},t) | 0 \rangle
\end{equation}
does not remain invariant. In the above equation $\Gamma$ is a space-time
independent matrix in spin, color, and flavor indices and $f(\vec{x})$ is
the relative wavefunction. A gauge invariant state can be constructed, however,
 by inserting a path-ordered gauge link operator between quark operators.
This amounts to including arbitrary numbers of soft gluons for transporting
color between quarks. Building on this idea, in a previous publication$^{(2)}$
we had
proposed a manifestly gauge invariant effective field theory describing the
interaction between heavy quarks, gluons, and quarkonium. The corrections
accruing from the incorporation of gauge symmetry turned out to be substantial
 for decay and fragmentation processes, as well as radiative transitions, and
this indicated the importance of a
correct treatment. However, the relation of the effective theory to QCD was
not transparent and it was not clear how the theory could be systematically
extended to higher orders.

In this paper we have developed a method which starts directly from QCD and
which does allow for a systematic treatment of all high momentum transfer
processes involving quarkonia such as inclusive decays, production, and
fragmentation. The natural expansion parameter is the quark relative velocity
which,
for a heavy quark system, is small. All of the non-perturbative physics turns
out to reside in a small number of matrix elements of gauge-invariant operators
 which are identified from symmetry considerations. The method used
 in this paper was inspired by the Feynman diagram treatment of deeply
inelastic scattering as originally developed by Ellis, Furmanski, and
Petronzio$^{(3)}$, and recently further expanded upon by Jaffe and Ji$^{(4)}$.

While this work was nearing completion, we received a preprint authored by
Bodwin, Braaten, and Lepage$^{(5)}$ which presents a comprehensive QCD analysis
of hard processes involving quarkonia. Their analysis is based upon a
non-relativistic formulation of QCD. We share similar conclusions although
these two approaches are quite different; with suitable identification of
parameters the results are identical. Perhaps an advantage of our
 method is its relative simplicity and its closer relation to the more
familiar relativistic QCD. On the other hand, the work of Bodwin et al$^{(5)}$
has
wider scope because it is ultimately aimed at also generating the static
properties
 of quarkonia through lattice calculations. We see the two approaches as
complementary to each other.\\
{\bf FORMALISM}

Our goal is to arrive at a systematically improvable, gauge-invariant,
description of all hard processes involving a $Q \bar{Q}$ system. By way of
introduction, consider the decay of a positive $\cal C$ parity state into 2
photons
 (Fig.1a) and the simplest Feynman graphs (Fig.1b) which contribute to it.
Decay widths of specific hadrons were computed from the leading approximation
 to these graphs long ago, and references may be found in Schuler$^{(1)}$. For
our purposes, it is useful to write the zero-gluon amplitude in Fig.1b as,
\begin{equation}
T_0^{\mu \nu} = \int \frac{d^4k}{(2 \pi)^4} Tr M(k) h^{\mu \nu}(k).
\end{equation}
$M(k)$ is the usual, but obviously non gauge-invariant, Bethe-Salpeter
amplitude,
\begin{equation}
M(k) = \int d^4x \;e^{i k.x} \langle 0 | T[\bar{\psi}(-x/2) \psi(x/2)] |
P\rangle.
\end{equation}

The tensor $h^{\mu \nu}(k)$ is the amplitude for two quarks, not necessarily
on their mass-shells, to annihilate into 2 photons. To leading order in
$\alpha_s$ this is,
\begin{equation}
h^{\mu \nu}(k) = -i e^2 \left [\gamma^\nu S_F(k+K) \gamma^\mu +\gamma^\mu
S_F(k-K) \gamma^\nu \right ].
\end{equation}
In Eqs 2-4, $x^\mu$ is the relative distance between quarks,
$k^\mu=\frac{1}{2}(p_1-p_2)^\mu$
is the quark relative momentum, $K^\mu=\frac{1}{2}(q_2-q_1)^\mu$ is the photon
 relative momentum and $S_F$ is the free fermion propagator. We shall refer
to $M(k)$ and $h^{\mu \nu}$ as $``$soft$"$ and $``$hard$"$ parts respectively
in
the following.

Now consider the fact that a large momentum of $O(m)$, where $m$ is the
heavy quark mass, flows through the single
 propagator in Fig.1b but that, on the other hand, the soft part has typical
quark momenta much less than $m$. This suggests that we expand the hard part in
 powers of $k^\alpha$,
\begin{equation}
h^{\mu \nu}(k) = \sum \frac{1}{n!} k^{\alpha_1}...k^{\alpha_n} U^{\mu
\nu}_{\alpha_1...\alpha_n},
\end{equation}
where,
\begin{equation}
U^{\mu \nu}_{\alpha_1...\alpha_n} = \frac{\partial}{\partial k^{\alpha_1}} ....
\frac{\partial}{\partial k^{\alpha_n}} h^{\mu \nu} |_{k=0}.
\end{equation}
Inserting Eq.5 into Eq.2 and integrating by parts gives for the amplitude,
\begin{equation}
T^{\mu \nu}_0(k) = Tr \sum_n {\cal M}^{\alpha_1...\alpha_n} U^{\mu
\nu}_{\alpha_1...\alpha_n},
\end{equation}
where,
\begin{equation}
{\cal M}^{\alpha_1...\alpha_n} = \frac{1}{n!} \langle 0 | \bar{\psi} \; i \!\!
\stackrel{\leftrightarrow}{\partial}^{\alpha_1}\!\!\!\!\!\!....\;i\!\!
\stackrel{\leftrightarrow}{\partial}^{\alpha_n} \!\!\!\!\! \psi | P \rangle.
\end{equation}
The matrix elements in Eq.8 have derivatives
$\stackrel{\leftrightarrow}{\partial}^\alpha = \frac{1}{2}
(\stackrel{\rightarrow}{{\partial}}^\alpha \!\!\!-\!\!\!
\stackrel{\leftarrow}{\partial}^\alpha)$
evaluated at zero relative quark separation. As we shall see  later, in a
non-relativistic model the $n=0$ matrix element is proportional to the
wavefunction at the origin, and so on. However, we do not need to appeal to
any particular model at this point.

Next consider the single gluon diagram in Fig.2. The corresponding amplitude
is,
\begin{equation}
T_1^{\mu \nu} = \int \frac{d^4k}{(2 \pi)^4} \frac{d^4k'}{(2 \pi)^4}   Tr \;
M_\rho (k,k') H^{\mu \nu \rho}(k,k').
\end{equation}
The $``$soft$"$ part $M_\rho (k,k')$ is a generalized B-S amplitude,
\begin{equation}
M_\rho (k,k') = \int d^4x d^4z \; e^{i k.x} e^{i k'.z} \langle 0 \left| T
\right[ \bar{\psi}(-x/2) A_\rho (z) \psi(x/2)\left] \right| P \rangle,
\end{equation}
and $A_\rho \equiv \frac{1}{2} \lambda^a A^a_\rho$ is the gluon field matrix.
The
$``$hard$"$ part $H^{\mu \nu \rho}$
is the annihilation amplitude for $\bar{Q} Q g \longrightarrow \gamma \gamma$.
 To leading order this is,\\
\begin{eqnarray}
 H^{\mu \nu \rho}(k,k') & = & -i e^2 g \left[ \gamma^\nu S_F(k+\frac{1}{2}
k'+K) \gamma^\rho S_F(k-\frac{1}{2} k'+K) \gamma^\mu \right]   \nonumber  \\
                       &   & + (\mu \leftrightarrow \nu).
\end{eqnarray}
Excluded are the diagrams such as in Fig.2b. These are properly included in
Fig.1b since the lines emerging from the blob correspond to interacting
fields. Expanding the hard part,
\begin{equation}
H^{\mu \nu \rho}(k,k') = \sum  \frac{1}{n! l!} k^{\alpha_1}...k^{\alpha_n}
k'^{\beta_1}...k'^{\beta_l} V^{\mu \nu
\rho}_{\alpha_1...\alpha_n,\beta_1...\beta_l},
\end{equation}
where,
\begin{equation}
V^{\mu \nu \rho}_{\alpha_1...\alpha_n,\beta_1....\beta_l} =
\frac{\partial}{\partial k^{\alpha_1}} .... \frac{\partial}{\partial
k^{\alpha_n}} \frac{\partial}{\partial k'^{\beta_1}} ....
\frac{\partial}{\partial k'^{\beta_l}} H^{\mu \nu \rho} |_{k=k'=0}.
\end{equation}
An integration by parts on $k, k'$ yields an alternate form for the amplitude
$T_1^{\mu \nu}$,
\begin{equation}
T_1^{\mu \nu}(k)=Tr \sum_{n l} M_\rho^{\alpha_1....\alpha_n,\beta_1....\beta_l}
V^{\mu \nu \rho}_{\alpha_1...\alpha_n,\beta_1...\beta_l},
\end{equation}
where,
\begin{equation}
M_\rho^{\alpha_1...\alpha_n,\beta_1...\beta_l} = \frac{1}{n! l!} \langle 0 |
\bar{\psi} \; i\! \stackrel{\leftrightarrow}{\partial}^{\alpha_1} \!\!\!\!\!
.... i\! \stackrel{\leftrightarrow}{\partial}^{\alpha_n} \!\!\!\!\! \psi \;i\:
\partial^{\beta_1} \!\
!\!\!
 ....i \: \partial^{\beta_l} A_\rho | P \rangle.
\end{equation}
The derivatives $i\! \stackrel{\leftrightarrow}{\partial}^{\alpha}$ act only
upon the quark operators.

Finally consider the two-gluon contribution to the amplitude shown in Fig.3.
The amplitude corresponding to Fig.3a is,
\begin{equation}
T_{2 a}^{\mu \nu} = \int \frac{d^4k}{(2 \pi)^4} \frac{d^4k'}{(2 \pi)^4}
\frac{d^4k''}{(2 \pi)^4}  Tr  M_{\rho' \rho''}(k,k',k'') H_a^{\mu \nu \rho'
\rho''} (k,k',k''),
\end{equation}
where,
\begin{eqnarray}
M_{\rho' \rho''}(k,k',k'') & = & \int d^4x d^4x' d^4x'' \;\;
e^{i(k.x+k'.x'+k''.x'')}  \nonumber \\
                           &   &  \;\;\;\;\; \langle 0 | \bar{\psi}(-x/2)
A_{\rho'}(x') A_{\rho''}(x'') \psi (x/2) | P \rangle.
\end{eqnarray}
The product of color matrix fields is appropriately symmetrised in the above
 because of Bose symmetry. The hard part $H^{\mu \nu \rho' \rho''}_a$ is, at
lowest order,
\begin{eqnarray}
H^{\mu \nu \rho' \rho''}_a \!\!(k,k',k'')\!\!\!\!\! & = &\!\!\!\!\! -i e^2 g^2
[ \gamma^\nu \! S_F(k\!+\!\!\frac{1}{2}\! k''\!\!+\!K) \! \gamma^{\rho''}\!\!
S_F(k\!+\!k'\!\!\!-\!\!\frac{1}{2}\! k''\!\!+\!K)\! \gamma^{\rho'} \!\!
S_F(k\!\!-
\!\frac{1}{2}\! k''\!\!+\!K) \gamma^\mu \nonumber \\
                                             &   &  \;\;\;\;\;\;\;\;\; +crossed
].
\end{eqnarray}
This may be expanded as before about $k=k'=k''=0$.
\vspace{.1in}

We now make the observation that the simple Ward identity,
\begin{equation}
\frac{\partial}{\partial p^\alpha} S_F(p) = -S_F(p) \gamma^\alpha S_F(p),
\end{equation}
leads to a number of useful relations. In particular,
\begin{equation}
V^{\mu \nu \rho} = -g U^{\mu \nu \rho},
\end{equation}
allows us to combine the $n=1$ term in Eq.7 and the $n=l=0$ term in Eq.14
into a gauge invariant sum,
\begin{equation}
{\cal M}^\alpha U^{\mu \nu}_\alpha+M^\alpha V^{\mu \nu}_\alpha = \langle 0 |
\bar{\psi} \;i\! \stackrel{\leftrightarrow}{D}^\alpha \!\!\!\! \psi | P \rangle
U^{\mu \nu}_\alpha.
\end{equation}
Similarly, the leading order term in the hard two-gluon amplitude is just the
 second order term in the zero-gluon amplitude,
\begin{equation}
H^{\mu \nu \rho' \rho''}(0) = g^2 U^{\mu \nu \rho' \rho''}.
\end{equation}
Collecting together appropriate terms leads to another gauge-invariant matrix
 element,
\begin{equation}
{\cal M}^{\alpha \alpha'} U^{\mu \nu}_{\alpha \alpha'} +
M^\alpha_\rho V^{\mu \nu \rho}_\alpha + M_{\rho' \rho''} H_a^{\mu \nu \rho'
\rho''}(0) = \frac{1}{2} \langle 0 | \bar{\psi}
i\!\stackrel{\leftrightarrow}{D}_{\rho'}
i\!\stackrel{\leftrightarrow}{D}_{\rho''} \! \! \psi | P \rangle U^{\mu \nu
\rho' \rho''}.
\end{equation}
Here $M_{\rho' \rho''}$ is just the leading order term from Eq.17. It is the
matrix element with all fields at the same space-time point,
\begin{equation}
M_{\rho' \rho''} = \langle 0 | \bar{\psi} A_{\rho'} A_{\rho''} \psi | P
\rangle.
\end{equation}
Next, look at the $n=0$ $l=1$ term in Eq.14:
\begin{equation}
\langle 0 | \bar{\psi} \psi i \partial_\beta A_\rho | P \rangle \partial'^\beta
H^{\mu \nu \rho} =\frac{i}{2}  \langle 0 | \bar{\psi} \psi ( \partial_\beta
A_\rho - \partial_\rho A_\beta) | P \rangle \partial'^\beta H^{\mu \nu \rho}
\end{equation}
The last step made use of
$\partial'^\beta H^{\mu \nu \rho} =- \partial'^\rho H^{\mu \nu \beta}$ ,
with all derivatives evaluated at $k=k'=0$. The quantity inside the brackets
is the abelian field strength tensor; the non-abelian part,
$i g \left[ A_\beta ,A_\rho \right]$, can be shown after some effort to arise
from the
amplitude in Fig.3b.

We now collect results together and summarize. The contribution from the 0,
1, 2 gluon diagrams in Figs.1-3 have been expanded in powers of relative
momentum, added together, and terms suitably rearranged. The total amplitude
is, neglecting higher powers of momentum,
\begin{eqnarray}
(T_0+T_1+T_2)^{\mu \nu} & = & Tr\left[ \; \langle 0 | \bar{\psi} \psi | P
\rangle h^{\mu \nu} + \langle 0 | \bar{\psi} i\!
\stackrel{\leftrightarrow}{D}_\alpha \psi | P \rangle \partial^\alpha h^{\mu
\nu}\right. \nonumber \\
                        &   &  \;\;\;\;\; + \langle 0 | \bar{\psi} i\!
\stackrel{\leftrightarrow}{D}_\alpha i\! \stackrel{\leftrightarrow}{D}_\beta
\psi | P \rangle \frac{1}{2} \partial^\alpha \partial^\beta h^{\mu \nu}
\nonumber \\
                        &   &  \left.\;\;\;\;\; + \langle 0 |  \bar{\psi}
F_{\alpha \beta} \psi | P \rangle  \frac{i}{2} \partial'^\alpha H^{\mu \nu
\beta} \right] .
\end{eqnarray}
The hard amplitude $h^{\mu \nu}(k)$, $H^{\mu \nu \alpha}(k,k')$, and their
derivatives are all evaluated at $k=k'=0$. We see that each term in the above
 is a product of a gauge invariant matrix element characteristic of the
decaying hadron and a simple, calculable, hard part. In the following, hadrons
 of specific $J^{PC}$ will be considered and the relative order of importance
 of the terms in Eq.26 will be explicated. Radiative corrections, which are
not included in the lowest order amplitudes $h$ and $H$, will be considered
separately.\\
\\
{\bf MATRIX ELEMENTS}

In the previous section the amplitude for a ${\cal C}$=+ quarkonium state to
decay
into 2 photons was expressed in terms of matrix elements of leading
gauge-invariant operators. Further progress requires we specify the angular
momentum and parity: we take J=0, ${\cal P}$=$-$ ($\eta_c$ and $ \eta_b$
mesons) for now,
 leaving other mesons for later analysis. From Lorentz invariance, and
invariance
under charge conjugation and parity, the only non-zero matrix elements are,
\begin{eqnarray}
\langle 0 | \bar{\psi} \psi | 0^{-+}\rangle & = & a_1 M^2 \gamma_5+a_2 M
P\!\!\!\!/ \gamma_5 \\
\langle 0 | \bar{\psi} i\! \stackrel{\leftrightarrow}{D}^\mu \!\!\!\! \psi |
0^{-+}\rangle & = & i b M^2 \sigma^{\mu \nu} \gamma_5 P_\nu \\
\langle 0 | \bar{\psi} F^{\mu \nu} \psi | 0^{-+}\rangle & = & c M^3
\epsilon^{\mu \nu \alpha \beta} \gamma_\alpha P_\beta \\
\langle 0 | \bar{\psi} i\! \stackrel{\leftrightarrow}{D}^\mu \!\!\!\! i\!
\stackrel{\leftrightarrow}{D}^\nu \!\!\!\! \psi | 0^{-+}\rangle & = &  M^2
\left[d_1 M^2 g^{\mu \nu}\!+\!d_2 P^\mu P^\nu\right] \gamma_5 +  \nonumber \\
                                             & & M^3 \!\! \left[e_1 g^{\mu \nu}
\! P^\alpha \!+\!  e_2 \! \frac{P^\alpha P^\mu P^\nu}{M^2}\!+\!e_3 (g^{\alpha
\mu} \! P^\nu \!+\! g^{\alpha \nu} \! P^\mu)\right] \!\!\!
 \gamma_\alpha \gamma_5
\end{eqnarray}
For brevity, color has not been explicitly indicated in Eqs 27-30. It is clear
 that in Eq.27 the two quarks must be in a color singlet and so, regarded as
 matrix in color space, only the unit operator appears on the right hand side.
 However, when one gluon appears, as in Eqs 28-29, the quarks may be in
either singlet or octet states and the corresponding constants
$b^{(1)}...b^{(8)}$ then appear on the RHS. In Eq.30, the two gluons can
combine into a color singlet or octet, and those in turn can combine with
the quark singlet and octet respectively to give an overall singlet. Clearly
this leads to a large number of constants which must be known in order to
describe corrections to $0^{-+}$ decay and if our approach is to have any
practical utility,
 this number must be curtailed according to some well defined principle.

To progress beyond this point, it will be necessary to specialize our
hitherto general discussion and select a particular gauge. The Coulomb gauge
is natural for this problem, as shown by vast experience with
positronium states. We shall not repeat here the arguments of Lepage
et al$^{(6)}$ who, using the QCD equations of motion in the Coulomb gauge,
make the following estimates,
\begin{eqnarray}
\partial^0 \sim m v^2  \;\;\;\;&,&\;\;\; g A^0 \sim m v^2    \nonumber \\
\vec{\partial} \sim m v  \;\;\;\;&,&\;\;\; g \vec{A} \sim m v^2 \nonumber \\
g \vec{E} \sim m^2 v^3   \;\;\;\;&,&\;\;\; g \vec{B} \sim m^2 v^4.
\end{eqnarray}
Here $v$ is the relative velocity of quarks - the small parameter in the
theory.
 The estimates (31) allow us to see that explicit gluons will not enter in
the leading order corrections to the naive quarkonium model. Therefore,
working to $O(v^2)$, one may effectively replace the covariant derivatives in
 Eqs 27-30 with ordinary derivatives, and ignore $\vec{E}$ and $\vec{B}$.

We next observe that tracing Eq.30 with $\gamma_5 \gamma_\mu$ or $\gamma_5
\gamma_\nu$,
 and using the equation of motion $iD\!\!\!\!/ \psi = m \psi$, yields the
constraint,
\begin{equation}
e_1+e_2+5 e_3 =0.
\end{equation}
Working in the rest frame of the meson $P^\mu = (M,\vec{0})$ and putting
$\mu=0 \;\; \nu=i$ yields $e_3 \sim O(v^3)$. Hence $e_1=-e_2 + O(v^3)$. Setting
$\mu =\nu =0$ yields $d_2=-d_1+O(v^3)$. This leaves us with having to deal
with $a_1, a_2, b, d_1$ and $e_1$ - five independent parameters at the
$O(v^2)$ level.

Further progress demands that we specialize a step further and specify a model
 framework for the $0^{-+}$ quarkonium state. We shall assume, in common with
 many other authors, that the Bethe-Salpeter equation with an instantaneous
kernel does provide an adequate description. This has been conveniently
reviewed by Keung and Muzinich$^{(7)}$ and we adopt their notation. The
momentum space B-S amplitude $\chi (p)$ satisfies the homogeneous equation,
\begin{equation}
\chi (p) = i G_0(P,p) \int \frac{d^4 p'}{(2 \pi)^4} \;\;K(P,p,p') \;\;\chi(p'),
\end{equation}
which, after making the instantaneous approximation
$K(P,p,p')=V(\vec{p},\vec{p'})$ and reduction to the non-relativistic limit
yields$^{(7)}$,
\begin{equation}
\chi (p) = \frac{M^{1/2} (M-2E) (E+m-\vec{p}.\vec{\gamma} )\gamma_5
(1-\gamma_0) (E+m-\vec{p}.\vec{\gamma}) \phi (| \vec{p} |)}{4 E (E+m)
( p^0+\frac{M}{2}-E+i\epsilon ) ( p^0-\frac{M}{2}+E-i\epsilon )}.
\end{equation}
The scalar wavefunction $\phi ( | \vec{p} |)$ is normalized to unity,
\begin{equation}
\int \frac{d^3p}{(2 \pi)^3} | \phi (| \vec{p} |) |^2 =1,
\end{equation}
and,
\begin{equation}
E = \sqrt{\vec{p}^2 +m^2}.
\end{equation}
Fourier transforming $\chi(p)$ to position space yields
$\langle 0 | \bar{\psi}(-x/2) \psi (x/2) | P\rangle$ from which, by tracing
with
appropriate gamma matrices, the coefficients $a_1, a_2, b, d_1$ and $e_1$ can
be
 extracted. So finally, to $O(v^2)$, one has a rather simple result,
\begin{eqnarray}
\langle 0 | \bar{\psi} \psi | 0^{-+}\rangle & = & \frac{1}{2}  M^{1/2} \phi (0)
\left(1+\frac{P\!\!\!\!/}{M}\right) \gamma_5+M^{-1/2} \frac{ \vec{\nabla}^2
\phi (0)}{M^2} P\!\!\!\!/ \gamma_5, \\
\langle 0 | \bar{\psi} iD^\mu \psi | 0^{-+}\rangle & = & \frac{1}{3} M^{1/2}
\frac{\vec{\nabla}^2 \phi (0)}{M^2} i \sigma^{\mu \nu} \gamma_5 P_\nu, \\
\langle 0 | \bar{\psi} F^{\mu \nu} \psi | 0^{-+}\rangle & = & 0, \\
\langle 0 | \bar{\psi} iD^\mu iD^\nu \psi | 0^{-+}\rangle & = & \frac{1}{6
}M^{5/2} \frac{\vec{\nabla}^2 \phi (0)}{M^2} \left(g^{\mu \nu} -\frac{P^\mu
P^\nu}{M^2}\right) \left(1+\frac{P\!\!\!\!/}{M}\right) \gamma_5.
\end{eqnarray}
Eqs.37-40 express hadronic matrix elements, upto $O(v^2)$, in terms of two
basic parameters: $\phi (0)$ and $\vec{\nabla}^2 \phi (0)$. These may be
obtained
 for any given phenomenological potential from a non-local Schr\"{o}dinger type
of equation$^{(7)}$.\\
\\
{\bf DECAY RATES}

The decay rate for $0^{-+} \longrightarrow 2 \gamma$ may be directly computed
 from Eqs.26 and 37-40. The calculation is facilitated by the observation
that, from invariance under time reversal, the crossed diagrams in Figs.1-3
exactly double the uncrossed ones. The result is,
\begin{equation}
(T_0+T_1+T_2)^{\mu \nu} = \frac{4 \sqrt{3}}{M^{3/2} (\frac{1}{4} M^2+m^2)}
\left( \phi (0) +\frac{8}{3} \frac{\vec{\nabla}^2 \phi (0)}{M^2} \right)
\epsilon^{\mu \nu \rho \lambda} q_{1 \rho} q_{2 \lambda} .
\end{equation}
The factor of $\sqrt{3}$ comes from the sum over colors. The quark mass $m$
differs from $M/2$ because of the strong binding,
\begin{equation}
\epsilon_B = 2 m - M.
\end{equation}
$\epsilon_B/M$ is of $O(v^2)$ from virial theorem, and thus of the
same order of magnitude as  $\nabla^2/M^2$. From Eq.41 it is simple to
get the decay rate (excluding radiative corrections),
\begin{equation}
\Gamma_{0^{-+} \rightarrow 2 \gamma} = \Gamma_0+\Gamma_B+\Gamma_C+\Gamma_R.
\end{equation}
In Eq.43, $\Gamma_0$ is the conventional result,
\begin{equation}
\Gamma_0 = \frac{12 \alpha_e^2 e_Q^4}{M^2} R^2(0),
\end{equation}
where $e_Q$ is the quark charge and $R(0)=\phi (0) \sqrt{4 \pi}$. $\Gamma_B$ is
the correction coming from $m \neq M/2$,
\begin{equation}
\Gamma_B = -2 \frac{\epsilon_B}{M} \Gamma_0,
\end{equation}
and $\Gamma_C$ is the term coming from differentiating the quark propagator
once, and then twice,
\begin{equation}
\Gamma_C = \frac{16}{3 M^2} \frac{\nabla^2 R(0)}{R(0)} \Gamma_0.
\end{equation}
Lowest order radiative corrections to $0^{-+} \longrightarrow 2 \gamma$ were
calculated by Barbieri et al$^{(8)}$ a long time ago. These are $O(v^2)$ too,
\begin{equation}
\Gamma_R = \frac{\alpha_s}{\pi} \left( \frac{\pi^2-20}{3}  \right) \Gamma_0.
\end{equation}
For decay into 2 gluons, the only difference in Eqs.44-46 is from the color
factor, but the 3 gluon vertex changes the form of the radiative correction,
\begin{equation}
\Gamma_{0^{-+} \rightarrow 2 g} = \frac{2 \alpha_s^2}{9 \alpha_e^2 e_Q^4}
(\Gamma_0+\Gamma_B+\Gamma_C+\Gamma'_R),
\end{equation}
where,
\begin{equation}
\Gamma'_R = \left (\beta_o \log \frac{\mu}{m} +\frac{159}{6}-\frac{31}{24}
\pi^2-11 \log 2+n_f (\frac{2}{3} \log 2-\frac{8}{9})\right )
\frac{\alpha_s}{\pi} \Gamma_0.
\end{equation}
The radiative corrections to the decay into gluons involves both the
renormalization scale $\mu$ and the renormalization scheme; for a discussion
of this point see Kwong et al$^{\; (9)}$.\\
\\
{\bf $1^{--}$ DECAY}

The formalism developed for two photon decay can be used quite trivially to
calculate the important decay $1^{--} \longrightarrow \gamma^* \longrightarrow
 l^+ l^-$. The $``$hard part$"$ is the single, momentum independent vertex,
$h^\mu = -i e \gamma^\mu$. There are therefore no corrections from expanding
the hard part, and the amplitude analogous to Eq.26 is simply,
\begin{equation}
T^\mu = Tr \langle 0 | \bar{\psi} \psi |P, \epsilon\rangle h^\mu,
\end{equation}
where $\epsilon^\mu$ is the vector meson polarization vector. Going to the
Coulomb gauge, and reducing the B-S equation, yields the amplitude analogous
to Eq.34 with the simple replacement
$\gamma_5 \longrightarrow \epsilon \!\!/$. Using ${\cal C}$ and ${\cal P}$
invariance of the matrix element in Eq.1, we find that to $O(v^2)$,
\begin{equation}
\langle 0 | \bar{\psi} \psi |P,\epsilon\rangle=\frac{1}{2} M^{1/2}
(1+\frac{\nabla^2}{M^2}) \phi(0) (1+\frac{\slashP}{M})
\slashepsilon-\frac{1}{2} M^{1/2} \frac{\nabla^2 \phi(0)}{3 M^2}
(1-\frac{\slashP}{M}) \slashepsilon
\end{equation}
This yields for the decay to leptons,
\begin{equation}
\Gamma_{1^{--} \rightarrow l^+ l^-} = \Gamma_{VW} +\Gamma_{rad}+\Gamma_{cor}.
\end{equation}
$\Gamma_{VW}$ is the usual Van Royen-Weisskopf$^{(10)}$ formula\footnote{The
lepton
mass correction is simply included by multiplying Eq.53 by $\sqrt{\!1\! -\!
\frac{m_l^2}{M^2}\!}
(\!1\! +\! \frac{2 m_l^2}{M^2}\!)$},
\begin{equation}
\Gamma_{VW} = \frac{4 \alpha_e^2 e_Q^2}{M^2} R^2(0),
\end{equation}
$\Gamma_{rad}$ is the radiative correction calculated some time ago by
Celmaster$^{(11)}$,
\begin{equation}
\Gamma_{rad} = -\frac{16}{3 \pi} \alpha_s \Gamma_{VW},
\end{equation}
and $\Gamma_{cor}$ is the correction term which comes from Eqs.50 and 51,
\begin{equation}
\Gamma_{cor} = \frac{4}{3 M^2} \frac{\nabla^2 R(0)}{R(0)} \Gamma_{VW}.
\end{equation}

Although we have used the same symbol $R(r)$ for the radial wavefunction of
the $1^{--}$ and $0^{-+}$ states, these wavefunctions are in principle
different. We shall return to this point later.\\
\\
{\bf COMPARISON}

In two important previous works, $O(v^2)$ corrections to $0^{-+}$ and
$1^{--}$ quarkonium decays have been evaluated. The first approach by Keung
and Muzinich$^{(7)}$ starts from the B-S equation with an instantaneous kernel
. Subsequently a non-relativistic reduction is made, followed by an expansion
 of the lowest order amplitude about the mass-shell value of the relative
momentum $\vec{p}^{\; 2}=(M/2)^2 \! \!-\! m^2=-m \epsilon_B$. The relevant
results of their
work are in Table 1. Their treatment does not satisfactorily resolve the issue
 of QCD gauge invariance of decay rates, although they do raise this question.

The second approach is that of Bodwin et al$^{(5)}$ which builds
systematically upon the rigorous formulation of non-relativistic QCD by
Lepage and co-workers$^{(6)}$. These authors introduce an ultra-violet cutoff
 $\Lambda$ of $O(m)$ and then construct a NRQCD Lagrangian by successively
adding new local interactions with 2-component fermion spinors. To account
for annihilation into photons, higher dimensional terms involving 4 fermion
operators are introduced into the Lagrangian, and their coefficients are
computed in a power series in $\alpha_s$ by putting the annihilating quarks
on mass-shell. The annihilation process, which cannot be got directly from
NRQCD, is taken into account via the optical theorem which relates
annihilation rates to the imaginary parts of $\bar{Q} Q \longrightarrow
\bar{Q} Q$ scattering amplitudes. Bodwin et al$^{(5)}$ express their results
(see Table 1) in terms of non-relativistic wavefunctions, their derivatives and
the quark mass $m$. They do not use the meson mass $M$. However, to enable a
comparison, we have expressed their results in terms of $M$ using
$\epsilon_B=2m-M$ after expanding to first order in $\epsilon_B/M$.
Note also that this definition of $\epsilon_B$ is opposite in sign to that
of Keung and Muzinich$^{(7)}$.

The third approach is that of this paper. For completeness we summarize this
too: the decay amplitude is given by the sum of all distinct Feynman diagrams
 leading from the initial quarkonium state to the final state. Each diagram is
 put into the form of a (multiple) loop integral with a kernel which is a
product of a hard part and a soft part. The hard part is treated with
perturbative QCD, and the soft part is analyzed into its different components
 with the use of Lorentz, $\cal C$, and $\cal P$ symmetries. Use of the QCD
equations of
motion enables separation of these components according to their importance
in powers of $v$. As the last step, a specific commitment to dynamics is made
and the B-S equation is used to express the components in the form of
wavefunctions.

The first comment regarding the results summarized in Table 1 is that all
six entries collapse into a single one, $1+\frac{4}{3} \frac{\nabla^2 R}{M^2
R}$, upon making
the identification\footnote{The relation between the binding energy and
$\frac{\nabla^2R}{R}$ is explained briefly as a renormalization
condition in Labelle et al$^{(12)}$ in the NRQED approach (see their Eqs.11 and
12). However, in
our case there is no principle which a priori constrains $\epsilon_B$ to bear
a fixed relation to $\frac{\nabla^2R}{R}$, and therefore both will be
considered
adjustable parameters. } $\epsilon_B/M=2 \nabla^2 R/M^2 R$. It is interesting
to
note that this condition is precisely that which follows for a potential
$V(\vec{r})$ which has $V(0)=0$. The Schr\"{o}dinger equation for this
potential at
$\vec{r}=0$ is ,
\begin{equation}
-\frac{2}{M} \nabla^2 R= -\epsilon_B R.
\end{equation}

However, it is well known that at small $r$
the potential is Coulomb-like, $V(r)\sim 1/r$. In this case the entries in
Table 1 are not identical for arbitrary choices of $\epsilon_B$ or
equivalently, the quark mass $m$. Furthermore, $\nabla^2 R$ is
apparently singular at the origin $\nabla^2 R(r) \sim  M R(r)/r$. As is clear
from
the uncertainty principle, the local kinetic energy becomes very large at
short distances and the expansion in powers of $v$ breaks down. This difficulty
may be circumvented by imagining that annihilation takes place in a diffused
region of size $O(1/m)$, i.e., that $R$ and $\nabla^2 R$ are quantities
renormalized
 at this scale. In any case, they are simply parameters which serve instead
of the parameters in Eqs.27-30.

In order to estimate the correction factors for charmonium, we used the
following
values of the independent parameters,
\bea
\alpha_s & = & 0.19 \nonumber \\
m & = & 1.43 \:\:GeV \nonumber \\
\frac{\nabla^2R}{R} & = & -0.7 \:\:GeV^2. \nonumber
\eea
With this particular choice of parameters and using Eqs.[43], [48] and [52],
the decay rates are calculated to be,
\begin{eqnarray}
\Gamma (J/\psi \longrightarrow e^+ e^-) & = & 5.61 \;\;KeV \nonumber \\
\Gamma (\eta_c \longrightarrow hadrons) & = & 9.99 \;\;MeV  \nonumber \\
\Gamma (\eta_c \longrightarrow 2 \gamma) & = & 6.48 \;\;KeV.
\end{eqnarray}
These values agree reasonably well with the experimentally measured decay
widths
which are$^{(13)}$,
\begin{eqnarray}
\Gamma (J/\psi \longrightarrow e^+ e^-) & = & 5.36 \pm 0.28 \;\;KeV \nonumber
\\
\Gamma (\eta_c \longrightarrow hadrons) & = & 10.3 \pm 3.6  \;\;MeV  \nonumber
\\
\Gamma (\eta_c \longrightarrow 2 \gamma) & = & 8.1 \pm 2.0 \;\;KeV.
\end{eqnarray}
In evaluating expressions (57), the radiative corrections are calculated at the
renormalization point $\mu=m^{(9)}$. The wavefunctions of $J/\psi$ and $\eta_c$
at the origin differ from each other to $O(v^2)$. This difference is neglected
in taking the ratio $\frac{\nabla^2R}{R}$. Their values are,
\bea
|R_{J/\psi}|^2 = 0.978 \;\; GeV^3 \nonumber \\
|R_{\eta_c}|^2  = 0.936 \;\; GeV^3. \nonumber
\eea

One remark concerns the value of $\alpha_s$ used above, which differs from the
value deduced from deep inelastic scattering, $\alpha_s(m_c) \approx 0.3$.
The reason is the following: the value of the parameter
$\frac{\nabla^2 R}{R}$ depends upon the value of $\alpha_s$ chosen and,
for smaller values of $\alpha_s$, this is  negative. The corresponding
 values of the wavefunctions of $J/\psi$ and $\eta_c$ at the origin should
differ from each other by $O(v^2)$ by the assumptions used in this paper.
However, for larger values of $\alpha_s$,
$\frac{\nabla^2 R}{R}$ becomes positive and correspondingly the difference
between the wavefunctions becomes  rather large. For example, for
$\alpha_s$=0.24, we have, $\frac{\nabla^2 R}{R}\!=\!2.8\: GeV^2 $, with
$|R_{J/\psi}|^2\!\! =\! 0.582 \; GeV^3$ and $|R_{\eta_c}|^2 \!\!  = \!0.194 \;
GeV^3$.
The difficulty in using large values of $\alpha_s$ has also been noted by
Consoli and Field$^{(13)}$, and suggests that $O(\alpha_s^2)$ radiative
corrections to charmonium decays may well be significant.

In conclusion, we have investigated higher order corrections to the decay of
$0^{-+}$
and $1^{--}$ heavy quarkonia and shown how these corrections can be
systematically
incorporated in terms of various bound state matrix elements of gauge-invariant
 quark and gluon operators. Investigation of $\cal P$=+, $\cal C$=+ states -
which correspond to
 P-waves in the n.r. limit - is in progress. We are also currently calculating,
using
 the framework developed in this paper, the more complicated case of the decay
of
 negative $\cal C$ parity quarkonium into 3 gluons/photons. This will enable a
more detailed
comparison of theory vs experiment.

\vspace{4.5cm}
\centerline{\bf Acknowledgments}

We wish to thank Xiangdong Ji for a fruitful discussion. This work was
supported in
parts by funds provided by NSF Grant No. INT-9122027. H.K gratefully
acknowledges
the financial support from the Akhter Ali Memorial Fellowship Fund for her
doctoral research .

\clearpage

\centerline{\bf Figure Captions}
\hspace{-1.2cm}
Figure:1 a) Heavy quarkonium decay into two photons. b) The lowest order
diagrams contributing to $Q \bar{Q} \longrightarrow 2 \gamma$.\\
Figure:2 a) Single gluon diagrams. b) Interacting quark field diagram
 which is properly included in Fig.1b.\\
Figure:3 a) Two gluon diagrams. b) Three gluon vertex diagrams.

\clearpage
\begin{table}
\centerline{ \begin{tabular}{|l|c|c|} \hline
&   $0^{-+} \longrightarrow  2 g$  &    $1^{--}
\longrightarrow  l^+ l^- $   \\ \hline
Keung et al$^{(7)}$   &  {\Large $1+\frac{2}{3} \frac{\epsilon_{\mbox{\tiny
B}}}{M}$ }
& {\Large   $1+\frac{2}{3} \frac{\epsilon_{\mbox{\tiny B}}}{M}$  }   \\ \hline
Bodwin et al$^{(5)}$ & {\Large $1-\frac{2 \epsilon_{\mbox{\tiny B}}}{M}
+\frac{16}{3 M^2}
\frac{\nabla^2 R}{R}$ }& {\Large $ 1-\frac{2 \epsilon_{\mbox{\tiny B}}}{M}
+\frac{16}{3 M^2}
\frac{\nabla^2 R}{R} $ }\\ \hline
This work             &  {\Large $1-\frac{2 \epsilon_{\mbox{\tiny B}}}{M}
+\frac{16}{3 M^2}
\frac{\nabla^2 R}{R}$ }& {\Large $1+\frac{4}{3 M^2} \frac{\nabla^2 R}{R}$}\\
\hline
\end{tabular}                                             }
\caption{{\em A comparison of the $O(v^2)$ correction factor, excluding
radiative corrections, which multiply the zeroth order formulae for the
electromagnetic decay of quarkonium states. M is the hadron mass,
$\epsilon_B = 2 m \!\!-\!\!M$ is the binding energy, and both $R$ and
$\nabla^2 R$ are evaluated at $r=0$. Note that all six entries become
identical upon making the identification $\epsilon_B/M=2 \nabla^2 R/M^2 R$. }}
\end{table}

\end{document}